# An Efficiency Firmware Verification Framework for Public Key Infrastructure with Smart Grid and Energy Storage System

Jhih-Zen Shih[1], Cheng-Che Chuang[2], Hong-Sheng Huang[3][0009-0000-8531-4196], Hsuan-Tung Chen[4] and Hung-Min Sun[5]

[1,5]Department of Computer Science, National Tsing Hua University, Hsinchu, Taiwan
s112062587@m112.nthu.edu.tw[1]
hmsun@cs.nthu.edu.tw[5]
[2,3]Institute of Information Security, National Tsing Hua University, Hsinchu, Taiwan
s112164519@m112.nthu.edu.tw[2]
ray.h.s.huang@m111.nthu.edu.tw[3]
[4]Information and Communications Research Laboratories, Industrial Technology Research Institute, Hsinchu, Taiwan
hsuantung@itri.org.tw[4]

**Abstract.** As a critical component of electrical energy infrastructure, the smart grid system has become indispensable to the energy sector. However, the rapid evolution of smart grids has attracted numerous nation-state actors seeking to disrupt the power infrastructure of adversarial nations. This development underscores the urgent need to establish secure mechanisms for firmware updates, with firmware signing and verification serving as pivotal elements in safeguarding system integrity. In this work, we propose a digital signing and verification framework grounded in Public Key Infrastructure (PKI), specifically tailored for resource-constrained devices such as smart meters. The framework utilizes the Concise Binary Object Representation (CBOR) and Object Signing and Encryption (COSE) formats to achieve efficient data encapsulation and robust security features. Our approach not only ensures the secure deployment of firmware updates against the convergence of information technology (IT) and operational technology (OT) attacks but also addresses performance bottlenecks stemming from device limitations, thereby enhancing the overall reliability and stability of the smart grid system.

## 1    Introduction

As global energy demand continues to rise, smart grids are gradually becoming the core of modern power systems, progressively replacing traditional grids and evolving into a critical infrastructure. On the other hand, traditional grids are no longer able to effectively address current challenges in terms of energy efficiency, maintenance costs, and environmental impact. The introduction of smart grids not only provides feasible solutions to these issues but also significantly enhances the flexibility and reliability of power systems. Nevertheless, the widespread adoption of smart grids also brings new challenges, particularly the threat from malicious attackers, especially concerning the security of system firmware updates. Some studies have proposed hardware-level solutions, such as enhancing protection against malicious code updates by limiting the update window [1]. However, these methods typically rely on hardware support, making them difficult to implement widely for resource-constrained devices or systems that cannot upgrade hardware. Therefore, our research focuses on software-level protection measures, using efficient data encapsulation and signing formats to address the security and performance bottlenecks in firmware updates, without altering the underlying hardware infrastructure.

In the past, the security of smart grids mainly relied on traditional signing technologies. Although these technologies are still effective in some scenarios and can ensure data integrity, they may not provide sufficient support for resource-constrained devices, such as smart meters and remote terminal units (RTUs). The limited computational power and storage capacity of these devices make the high computational demands and large data sizes of traditional signing formats a performance bottleneck and a potential security risk. Therefore, selecting an efficient signing format that can ensure security while accommodating the limitations of the devices has become a major challenge in the field of smart grid security.

To address this need, we propose a digital signature and verification framework based on the Public Key Infrastructure (PKI) model. This framework adopts the Concise Binary Object Representation (CBOR) [2] and COSE formats, aiming to efficiently handle data encapsulation and signature mechanisms. The CBOR format is highly compact, making it particularly suitable for devices with limited computational power and storage capacity. Meanwhile, COSE provides an efficient digital signature mechanism that significantly reduces resource consumption without compromising security.

The core purpose of selecting these technologies is to address the performance bottlenecks introduced by traditional signature formats while ensuring the integrity and security of system data. Among these, the CBOR Object Signing and Encryption (COSE) technology provides an efficient and lightweight solution [3] , making it particularly suitable for applications in resource-constrained devices. Next, we will detail how these technologies are integrated into our framework and their role in the firmware update architecture of smart grids, enhancing the overall security, reliability, and stability of the system. The application of these technologies not only resolves the bottlenecks of traditional signature formats but also offers effective support for operational challenges encountered in the functioning of smart grids.

With the increasing number of devices in smart grids, issues related to coordination and synchronization have become more pronounced. In environments where tens of thousands of smart devices operate simultaneously, any delay or anomaly can lead to system instability. Traditional data encapsulation formats, due to their complex structures and high overhead, struggle to meet the needs of large-scale device coordination, especially when the system requires low-latency data transmission.

Moreover, according to the security reports [4][5][6][7] on smart grids, several well-known Advanced Persistent Threat (APT) groups, such as INDUSTROYER2, COSMICENERGY, and Fancy Bear, have been utilizing malware—including STUXNET, CaddyWiper, TRITON, and VPNFilter—to target Industrial Control Systems (ICS), as illustrated in Figure 1. Upon successfully infiltrating these systems, attackers gain control over critical components, such as Remote Terminal Units (RTUs), enabling them to manipulate the operational state of energy systems. These actions can result in intentional power outages, exemplified by the attack on Ukraine's power infrastructure, allegedly conducted by state-level actors from Russia. Additionally, the United States Department of Homeland Security has reported analogous attacks on the nation's energy infrastructure, exploiting vulnerabilities within Supervisory Control and Data Acquisition (SCADA) systems.

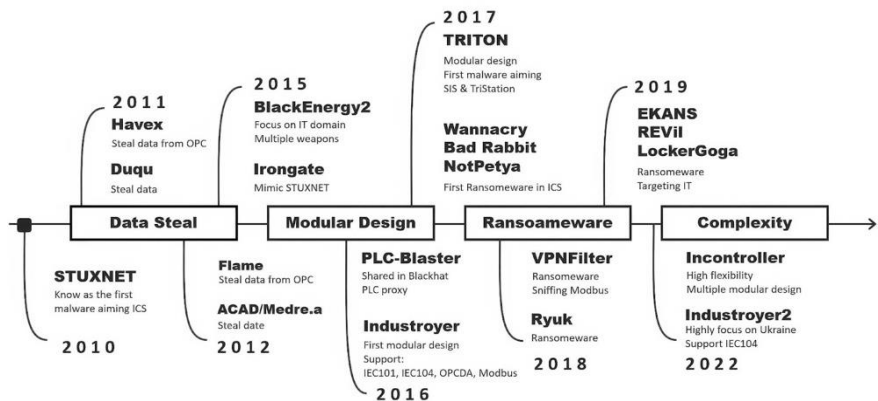

**Fig.1** Overview of ICS Malware and APT Groups from 2010 to 2022.

Against these backdrops, our framework adopts the COSE (CBOR Object Signing and Encryption) format to address operational challenges in smart grids. By leveraging the compact CBOR encoding, COSE significantly reduces data overhead and transmission time, thereby enabling real-time responses. This approach effectively meets the requirements of resource-constrained devices while ensuring reliability and security in data processing. The efficient encapsulation scheme resolves bottlenecks in device synchronization, enhancing both system stability and responsiveness. Additionally, our framework mitigates the risk of malware propagation into firmware. The subsequent sections will illustrate the application of these techniques in firmware updates, providing experimental validation of their performance and benefits, thereby offering robust support for the practical deployment of smart grids.

## 2    Methods

### 2.1    System Workflow

Our system workflow aims to provide a secure and reliable firmware update mechanism for devices such as smart meters. When a device receives a firmware update file from the vendor, the file must first undergo verification to ensure its source's credibility and the integrity of the file. The verification process will be based on digital signatures, hash values, and other information to ensure that the firmware update process is not compromised or tampered with. Only when the file is successfully verified will the system proceed with subsequent firmware updates or other instructions, ensuring the security and stability of the update process.

### 2.2    Firmware Signing Process

During the firmware update process, the firmware developer uploads the new version of the firmware update file to the signing website. After being reviewed and approved by the administrator, the file will undergo the signing process. As shown in Figure 2, during the signing process, the administrator will sign the file using the corresponding public and private key certificates based on the selected signing algorithm. Once the signing is complete, the packaged COSE file will be provided for the uploader to download. When devices such as smart meters receive the packaged file, they will verify the file using the validation mechanism to ensure that it has not been tampered with and that its source is trustworthy.

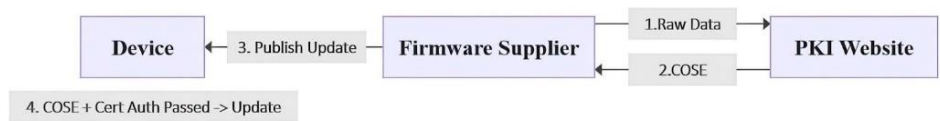

**Fig. 2** Firmware Signing and Distribution Process

### 2.3    Pre-Signature Review Mechanism

In our architecture, when the PKI signing entity performs the signing, the firmware vendor must upload the complete update file, and the PKI signing entity determines whether to proceed with the signing. To enhance security, a pre-signing review mechanism can be added, which performs multiple checks on the file, including source code scanning, static analysis, and threat assessment. Only when the file successfully passes these tests and meets the predefined security standards can the signing process proceed, thereby preventing various attack threats such as Supply Chain Attacks and Malware Injection. This mechanism not only strengthens the system's defense against untrusted files but also improves the transparency and trustworthiness of the firmware supply chain, fundamentally ensuring the security and reliability of the entire update process.

## 2.4 COSE Data Structure Design

COSE (CBOR Object Signing and Encryption) is based on the CBOR format and defined according to RFC 8152. It is designed to support security functions such as signing, encryption, and authentication. As illustrated in Figure 3 [5], the COSE data structure is encapsulated into four distinct blocks: Protected Header, Unprotected Header, Payload, and Signature, each serving a specific function. The Protected Header block contains critical metadata required for encryption, such as encryption algorithm identifiers and key management protocols. This information is encrypted to ensure the integrity and confidentiality of the message. In contrast, the Unprotected Header block contains non-encrypted metadata, such as message identifiers and timestamps. While this part is not encrypted, providing greater flexibility, it requires additional security measures to prevent the leakage of sensitive information. The Payload block is the core of the COSE message, containing the actual data being transmitted. Depending on the requirements, this section can optionally be encrypted to protect the data from unauthorized access or tampering. Finally, the Signature block is used to verify the integrity of the message and the identity of the sender by signing the data with a private key, ensuring that the receiver can validate the authenticity and source of the message. The collaborative design of these blocks ensures a balance between security, efficiency, and flexibility in the COSE data structure.

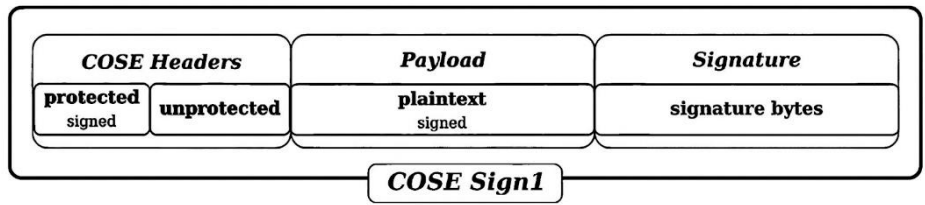

**Fig. 3** Schematic diagram of the COSE structure [8]

## 2.5 Verification Process and Algorithm

Figure 4 illustrates the secure and trusted process for authenticating message packets in our system. The COSE verification procedure first parses the received CBOR-formatted message and extracts the certificate chain from the x5chain. If the chain contains multiple certificates, the system will verify the certificate chain to ensure that the trust relationship between the issuer and signer of each certificate conforms to the specifications. Next, the system uses the pre-stored issuer certificate to validate the COSE message's signature. If the verification fails, the system will switch to the default certificate set for validation. Once the certificate verification is successful, the system uses the public key of the signing certificate to verify the entire COSE message. If all steps pass, the system will extract and store the payload for subsequent operations, such as firmware updates. If any verification step fails, the signed packet will be considered invalid, and an error message will be returned.

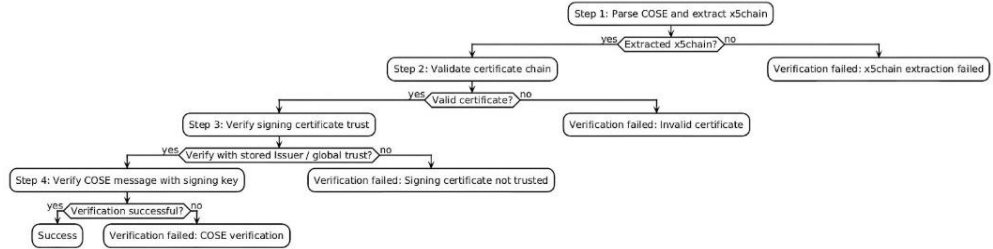

**Fig. 4** Firmware Signature COSE Verification Flowchart

## 3 Results

### 3.1 COSE Packet Demonstration

To address the security challenges faced by resource-constrained devices in smart grids, we designed a digital signing and verification framework based on the COSE format. Figure 5 presents an example of a COSE packet from our design, illustrating its four main components: the protected header, unprotected header, payload, and signature. Each component serves a specific function, ensuring the integrity and reliability of the packet during transmission.

**Protected Header:** This section contains critical information such as the signing algorithm and timestamp label. As these elements are essential for verification and must remain unaltered, they are included in this section and hashed together with the payload. This design ensures that any tampering during transmission can be detected, thereby safeguarding the integrity of the message.

**Unprotected Header:** This section is excluded from the hashing process and primarily includes the x5chain label, which stores the signer's certificate chain. We designed the framework to convert certificates into CBOR format and embed them within the message, allowing recipients to directly extract and use them without requiring additional transmissions. This saves bandwidth and improves efficiency. Since certificates inherently carry their own signatures for validity verification, they are placed in this section without participating in the overall hashing process. This design enables recipients to verify the authenticity of the signer and the message source, completing the verification with a single packet without needing additional certificate files.

**Payload:** This section contains the actual message to be authenticated and serves as the core of the COSE packet. In our framework, the payload includes firmware update data and version information, along with a hash value of the data to facilitate quick integrity verification.

**Signature:** Finally, the signature section contains the signature generated by encrypting the hash result of the protected header and payload with the signer's private key. This ensures the integrity of the packet's content, and only recipients with the corresponding public key can verify the authenticity and validity of the signature.

```
{
  "protected_header": {
    "algorithm": "selected algorithm"
  },
  "unprotected_header": {
    "timestamp": "<Timestamp>",
    "cert_chain": [
      "<cbor certificate  1>",
      "<cbor certificate  2>"
    ]
  },
  "payload": "<payload data>",
  "signature": "<encrypted signature>"
}
```

**Fig. 5** COSE Packet Example

### 3.2    Verification Process Demonstration

To validate the feasibility of the proposed COSE-based digital signature verification framework, we conducted experiments to evaluate the device's capability to parse and verify firmware update packets upon receipt. In the tests, we designed and implemented a verification program on the device side, which can perform a complete signature verification process solely using the firmware update packet itself and the pre-stored Issuer certificate in the device.

During the verification process, the device first receives the firmware update packet, then parses its COSE structure and utilizes the signature and the signing algorithm specified in the protected header for verification. The verification relies only on the pre-stored Issuer certificate on the device, without requiring additional network requests or extra verification credentials, successfully determining the authenticity and integrity of the update file. The experimental results indicate that, in multiple tests, the device accurately and consistently completed the verification process, promptly rejecting the update when an invalid or tampered signature was detected.

### 3.3    Comparison of COSE and Traditional PKI Message Encapsulation Sizes

To evaluate the performance of the COSE-based digital signing and verification framework on resource-constrained devices, we compared it with traditional PKI formats. In the experiment, a firmware update message containing a signature and certificate chain was selected as the test sample. The experimental group utilized the COSE format, while the control group employed traditional PKI encapsulation using DER-encoded certificates and JSON formatting. To ensure a fair comparison, all encapsulation formats included identical data content. Table 1 presents a comparison of the encapsulation sizes between the COSE and JSON formats for a fixed payload size. Under the condition where both formats use the ECDSA_SHA256 algorithm for signing and a zero-byte payload, the advantages of CBOR over JSON are evident, particularly in certificate storage and message structure representation. The variable in this comparison is the number of certificates included in the certificate chain. Experimental data demonstrate that CBOR-based storage achieves approximately 30% savings in storage space compared to JSON.

**Table 1.** Comparison of COSE and JSON Packet Sizes with Fixed Payload Size

| Payload size | | COSE message size | | JSON message size | | Reduced ratio |
|---|---|---|---|---|---|---|
| 1 | Byte | 1038 | Bytes | 1497 | Bytes | 0.693 |
| 1 | KB | 2063 | Bytes | 2861 | Bytes | 0.721 |
| 100 | KB | 101 | KB | 135 | KB | 0.749 |
| 1 | MB | 1025 | KB | 1366 | KB | 0.750 |
| 100 | MB | 100 | MB | 133 | MB | 0.750 |

To further investigate the impact of different payload sizes on encapsulation efficiency, we adjusted the payload size and compared the performance of the COSE and JSON formats under varying payload conditions. The data was signed using the ECDSA_SHA256 algorithm, with the certificate chain containing two certificates. The comparison results are shown in Table 2. The experimental data reveal that as the payload size increases, the signing overhead for the COSE framework remains nearly constant. In contrast, the encapsulation size of the traditional PKI format is significantly larger than that of the COSE format. This disparity is primarily due to the inflation effect of JSON when handling binary data. JSON requires binary data to be converted into Base64 encoding, increasing the data size by approximately 33%. Additionally, JSON contains numerous key names and symbols, further inflating the encapsulation size. In comparison, the COSE format, utilizing the compact CBOR encoding, can process binary data directly, avoiding these overheads and demonstrating higher compression efficiency and smaller encapsulation sizes. For firmware files of a certain size, the space savings approach 25%.

**Table 2.** Comparison of COSE and JSON Packet Sizes with Variable Payload Size

| Cert count | COSE message size | | JSON message size | | Reduced ratio |
|---|---|---|---|---|---|
| 1 | 585 | Bytes | 882 | Bytes | 0.663 |
| 2 | 1037 | Bytes | 1493 | Bytes | 0.695 |
| 3 | 1491 | Bytes | 2108 | Bytes | 0.707 |

The experimental results confirm that the COSE format shows significant potential for applications in resource-constrained environments such as smart grids. It can markedly reduce data transmission overhead and alleviate device processing burdens while maintaining security and data integrity.

# 4     Conclusion

As a critical part of power infrastructure, smart grids face threats from nation-state attackers, especially in the context of the convergence of information technology (IT) and operational technology (OT) attacks, making security challenges increasingly complex. In this environment, the security requirements for firmware updates are becoming more urgent, and firmware signing and verification have become core measures to prevent cyberattacks and unauthorized tampering. For resource-constrained devices, how to efficiently implement signature verification has become a major challenge. The limitations in computational power, storage capacity, and network bandwidth make designing lightweight security frameworks a critical issue that needs to be addressed.

In this work, we successfully propose a digital signature and verification framework based on a public key infrastructure (PKI) model, specifically designed for resource-constrained devices, and combined with CBOR and COSE formats to address security challenges in firmware updates. The framework is designed to balance high security with efficiency and resource consumption. This design effectively addresses the challenges brought by device performance bottlenecks, improving the operational efficiency of resource-constrained devices through efficient data processing and a simplified serialization process.

Our framework lies in the application of the CBOR format, which reduces the computational load of serialization and deserialization by using compact binary representations of data, shortening processing times and reducing power consumption. This is particularly important for resource-constrained devices that need to operate for extended periods in a smart grid environment, as it not only improves operational efficiency but also extends device lifespan. Additionally, embedding the signing certificate within the COSE message avoids the need for the device to fetch certificates externally, thus saving network bandwidth and improving efficiency.

This framework based on CBOR and COSE successfully addresses the security challenges of resource-constrained devices and achieves efficient data processing and robust security. These designs enhance the overall reliability of the smart grid system, strengthening its ability to resist external attacks and unauthorized operations, ensuring system stability and long-term operational reliability. Additionally, the COSE format's certificate chain and signature verification functions further bolster the trustworthiness of firmware update data, effectively preventing unauthorized updates from untrusted sources and enhancing the framework's ability to respond to potential network attacks within the smart grid environment, thereby ensuring data integrity and system stability.

We demonstrate the CBOR and COSE-based digital signature and verification framework, which can provide an efficient and secure firmware update mechanism for resource-constrained devices, enhancing the overall system security to address increasingly complex security challenges.